\documentclass[11pt]{article}
\usepackage{latexsym}
\usepackage{mathrsfs}
\usepackage{amsmath,amsthm}
\usepackage{graphicx,epstopdf,epsfig,multirow,epic,bm}
\usepackage{color,fancyvrb}
\usepackage{colortbl}
\usepackage{multicol}
\usepackage{amssymb}

\usepackage{makeidx}
\usepackage{showidx}
\usepackage{textcomp}
\usepackage{paralist}

\theoremstyle{plain}
\newtheorem{thm}{Theorem}

\theoremstyle{definition}

\theoremstyle{plain}

\theoremstyle{problem}

\theoremstyle{plain}

\theoremstyle{plain}

\theoremstyle{plain}

\DeclareGraphicsExtensions{.eps,.eps.gz}
\normalsize \rm
\newcommand{\js}{\hfill $\square$}
\newcommand{\ud}{\mathrm{deg}}

\begin{document}

\thispagestyle{empty}

\begin{center}
{\Large \bf Growing Network Models Having\\[12pt] Part Edges Removed/added Randomly}\\[12pt]
\end{center}

\begin{center}
{\large Bing \textsc{Yao}$^{a,}$\footnote{Corresponding author, Email: yybb918@163.com}\quad Xiaomin \textsc{Wang} $^{a}$ \quad Xia \textsc{Liu} $^{a}$  \quad Jin \textsc{Xu} $^{b}$}\\[8pt]
{\footnotesize a. College of Mathematics and Statistics, Northwest
Normal University, Lanzhou, Gansu 730070, China\\
b. School of Electronics Engineering and Computer Science, Peking University, Beijing,
100871, China}\\[12pt]
\end{center}
\begin{quote}
\textbf{Abstract:} Since network motifs are an important  property of networks and some networks have the behaviors of rewiring or reducing or adding edges between old vertices before new vertices entering the networks,  we construct our non-randomized model $N(t)$ and randomized model $N'(t)$ that have the predicated fixed subgraphs like motifs and satisfy both properties of growth and preferential attachment by means of the recursive algorithm from the lower levels of the so-called bound growing network models. To show the scale-free property of the randomized model $N'(t)$, we design a new method,  called edge-cumulative distribution, and democrat two edge-cumulative distributions of $N(t)$ and $N'(t)$ are equivalent to each other.\\
\textbf{Keywords:} network motifs; growing network; distribution; scale-free networks.\\
\textbf{PACS numbers:} \textsf{89.75.Da}; \textsf{02.10.Ox}; \textsf{89.75.Fb}
\end{quote}

\vskip 0.6cm

\section{Introduction and concepts}

Barabasi and Albert \cite{Barabasi-Albert1999} observed that both ingredients of \emph{growth and preferential attachment} are needed for the development of the stationary power-law distribution, since growth and preferential attachment are \emph{mechanisms} common to a number of complex systems, including business networks, social networks (describing individuals or organizations), transportation networks, and so on.
Directed networks have been discussed in \cite{Barabasi-Oltvai2004}, \cite{Dorogovtsev-Mendes-Samukhin2001}, \cite{Zhang-van-Moorsel2009} and \cite{Schwartz-Cohen-ben-Avraham-Barabasi-Havlin2002}. The authors in \cite{Comellas-Sampels2002} show vertex replacement and vertex addition methods to produce small-world and scale-free networks from a low diameter ``backbone'' network; specific recursive scale-free constructions with fixed degree distributions were investigated in \cite{Lu-Su-Guo2013} and \cite{Comellas-Fertin-Raspaud2004}.

Network motifs are an important local property of networks, and have been identified in a wide range of networks
across many scientific disciplines and are suggested to be the basic
building blocks of most complex networks \cite{Jiang-Tu-Chen-Su2006,Radu-Qasim-Barabasi-Oltvai2004}. Motifs are of notable importance largely because they may reflect functional properties, and they have recently gathered much attention as a useful concept to uncover structural design principles of complex networks, and may provide a deep insight into the network's functional abilities \cite{Masoudi-Nejad-Schreiber-Razaghi2012}. If the probability of a given subgraph to appear at least the same number of times as in the real network is smaller than a given threshold, the subgraph is considered a motif of the network \cite{Costa-Rodrigues-Travieso-Villas-Boas2006}. In \cite{Milo-Shen-Orr-Itzkovitz-Kashtan-Chklovskii-Alon2002, Shen-Orr-Milo-Mangan-Alon2002, Milo-Itzkovitz-Kashtan-Levitt-Shen-Orr-Ayzenshtat-Sheffer-Alon2004} the authors defined ``network motifs'' patterns of interconnections occurring in complex networks at numbers that are significantly higher than those in randomized networks, and they found such motifs in networks from biochemistry, neurobiology, ecology, and engineering.

We are motivated from the Barab\'{a}si-Albert (BA) model that is an algorithm for generating random scale-free networks using a preferential attachment mechanism, and motivated from a phenomenon that some groups of new vertices enter simultaneously into a network, not one by one. So, we will show recursive construction methods to generate models that are graphs having no multiple edges between the same pair of vertices and no self-edges that connects the same vertex at both ends, and some edges of the models can be removed randomly, and then some new edges can be added to the remainder randomly. For the purpose of simulation, our deterministic models are visual and satisfy both properties of growth and preferential attachment, and furthermore the construction of our models can be shown as the recursive scale-free algorithm from the lower levels of the so-called bound growing network models. Our models differ from those models in  \cite{Comellas-Fertin-Raspaud2004, Lu-Su-Guo2013, Zhang-Comellas-Fertin-Raspaud-Rong-Zhou2008, Zhang-Zhou-Su-Zou-Guan-2008, Zhang-Zhou-Fang-Guan-Zhang-2007,Zhang-Rong-Guo2006}, since we grow them by adding some graphs called \emph{seeds} and by making edges removed/added randomly. We use these seeds as motifs appeared in many hierarchical and biological networks. But these seeds are not directed in this paper.

As reported, some networks have the behaviors of rewiring or reducing edges between old vertices before new vertices entering the networks. Based on this observation, we have developed our evolutionary models that we will investigate in this article: one is called the \emph{uniformly $(r,F)$-growing network models} (abbreviation as $(r,F)$-ugnms) that have no edges removed/added; the another one is called the \emph{randomized $e$-bound growing network model} (abbreviation as randomized EBGN-model). Here, our ``$e$-bound growing'' is the same growing way as that shown in \cite{Zhang-Rong-Guo2006}, and is local not universal as \cite{Comellas-Fertin-Raspaud2004}. Our first step is to generate randomized EBGN-models. The generative method for obtaining a high level randomized EBGN-model is based on two phrases: (1) add new graphs like motifs to the so-called bound-edges of a lower level randomized EBGN-model; (2) the randomized mechanism: based on a probability $p_r$ remove some edges from this lower level randomized EBGN-model, and then add new edges to the remainder by another probability $p_a$. Our second step is to show the scale-free property of the randomized EBGN-models by means of the topological structure of $(r,F)$-ugnms. The term ``removing edges'' mentioned in this paper means that we delete edges and keep their ends in networks; and the term ``adding edges'' means that we join two vertices in networks by an edge if there is no edge between them. Clearly, it is not easy to obtain the degree spectrum of a randomized EBGN-model. As an alternative approach, we defined the edge-cumulative distribution for proving the scale-free formation of the randomized EBGN-models.

\section{Models having edges removed/added randomly}

Our randomized EBGN-model $N'(t)$ having edges removed/added will be constructed by the following deterministic algorithm-I. For short writing we use ``$i\in [m,n]$'' rather than ``$i=m,m+1,\dots ,n$'', where integers $n>m\geq l$; the number of elements of a set $X$ is denoted as $|X|$; the length of a path $P$ is denoted as $|P|$ in the whole paper. An edge $uv$ has its own ends $u$ and $v$. Let $F$ be a \emph{seed set} of finite connected graphs having $m_v~(\geq 1)$ vertices and $m_e~(\geq 0)$ edges, and let every connected graph $G\in F$ have its vertices $x_1, x_2, \cdots, x_{m_v}$. Here, the graphs in $F$ play the role of motifs or communities appeared in many real networks. We take an integer $r$ that holds $m_v\geq r\geq 1$, and call $r$ the \emph{bound thickness} as wall as every $G\in F$ a \emph{seed}. Let $N'_{v,t}$, $N'_{e,t}$ and $N'_{be}(t)$ be the numbers of vertices, edges and bound-edges of the randomized EBGN-model $N'(t)$, respectively.
\begin{verse}
\textbf{Algorithm-I}

\textbf{Step 1.} (Initialization) The initial model $N'(0)$ has no multiple edges and loops, and is connected. $N'(0)$ has its own vertex-set $V'(0)$ with $N_{v,0}=|V'(0)|\geq 2$ vertices and its own edge-set $E'(0)$ with $N'_{e,0}=|E'(0)|\geq 1$ edges. We define every edge $uv$ of $N'(0)$ as a \emph{bound-edge}. Let $B'(0)$ be the bound-edge set of $N'(0)$. Clearly, $B'(0)=E'(0)$. For $t=1$, the new model $N'(1)$ can be obtained by doing the following operation-I.\\
\textbf{Operation-I:} Add a \emph{seed} $G\in F$ to each bound-edge $uv$ of $N'(0)$, and join every vertex $x_i$ of $G$ with vertex $u$ and vertex $v$ to form two edges $x_iu$ and $x_iv$, respectively; and select randomly vertices $x_{i_1},x_{i_2},\dots, x_{i_r}$ to define $2r$ edges $x_{i_j}u$ and $vx_{i_j}$ as the bound-edges for $j\in [1, r]$.

\textbf{Step 2.} (Iteration including growth and removing/adding edges) For $t\geq 2$, $N'(t)$ is obtained by doing the operation-I to each bound-edge of $N'(t-1)$; remove randomly some edges from $N'(t-1)$ by the probability $p_r$, and the remainder is denoted as $T'(t-1)$; and then add randomly new edges to $T'(t-1)$ according to the probability $p_a$, where $0<p_r,p_a<1$ and $p_r$ is independent of $p_a$.
\end{verse}

\begin{figure}[h]
\centering
\includegraphics[height=6cm]{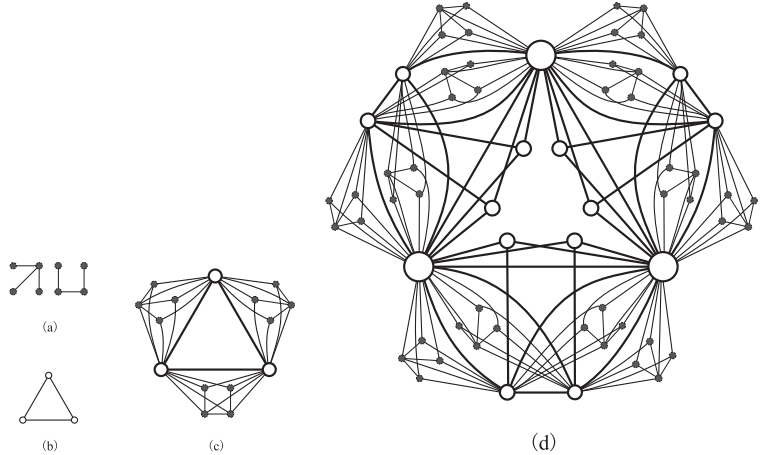}
\caption{\label{fig:randomized-example-1} {\small (a) Two seeds with $m_v=4$ and $m_e=3$; (b) $N'(0)$; (c) $N'(1)$ having $15$ vertices and $36$ edges; (d) the procedure of growing $N'(1)$ having $63$ vertices and $188$ edges by the bound thickness $r=2$.}}
\end{figure}

\begin{figure}[h]
\centering
\includegraphics[height=5.8cm]{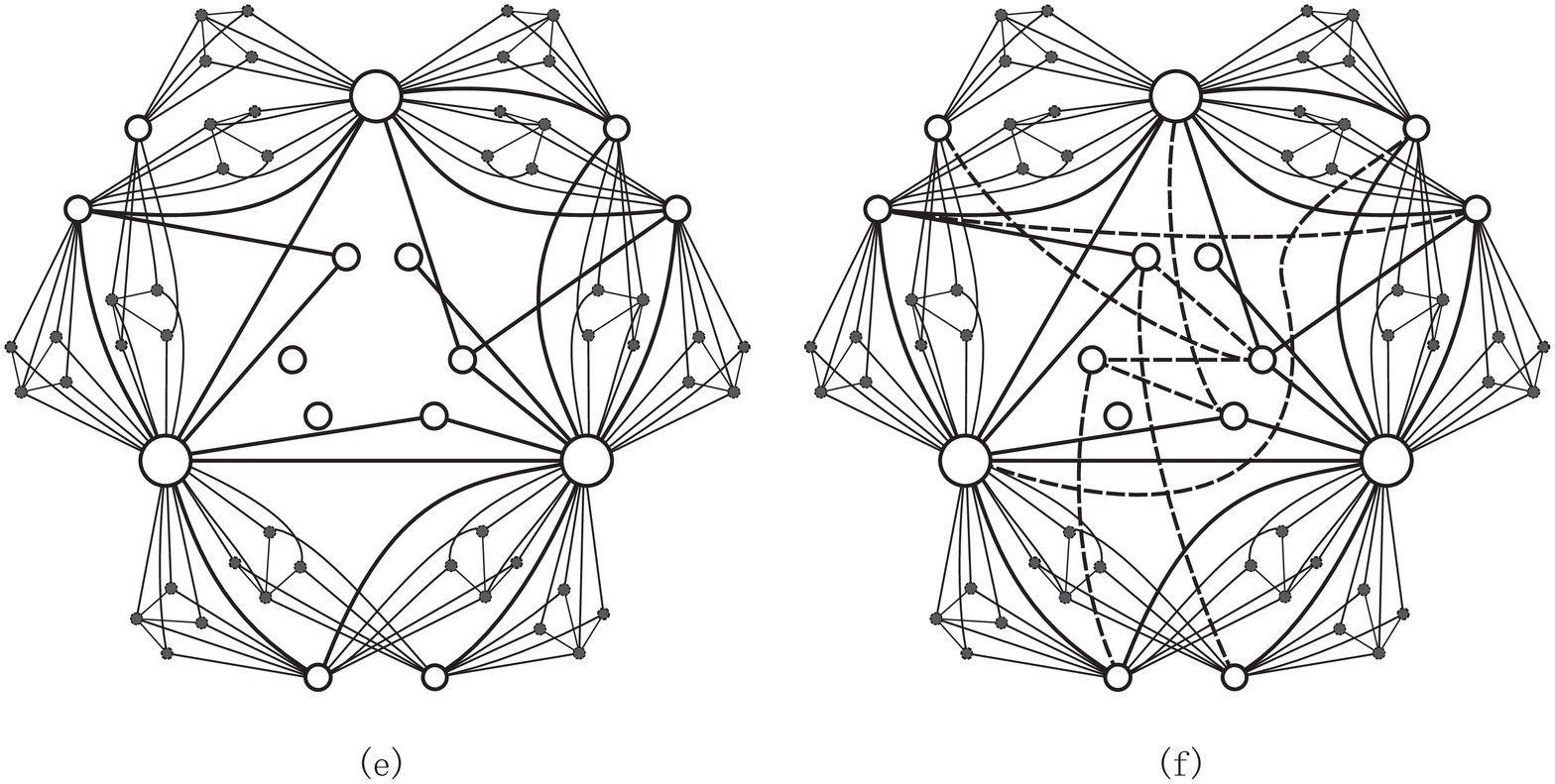}
\caption{\label{fig:randomized-example-2} {\small (e) $T'(1)$ is obtained by removing $18$ bold-edges of $N'(1)$ shown in Fig.\ref{fig:randomized-example-1} by the probability $p_r=\frac{1}{2}$; (f) the randomized EBGN-model $N'(2)$ obtained by adding $9$ new dashing-edges to $T'(1)$ by the probability $p_a=\frac{1}{2}$. Here, the number of edges of $N'(2)$ is less than that of $N'(1)$.}}
\end{figure}

\begin{figure}[h]
\centering
\includegraphics[height=6cm]{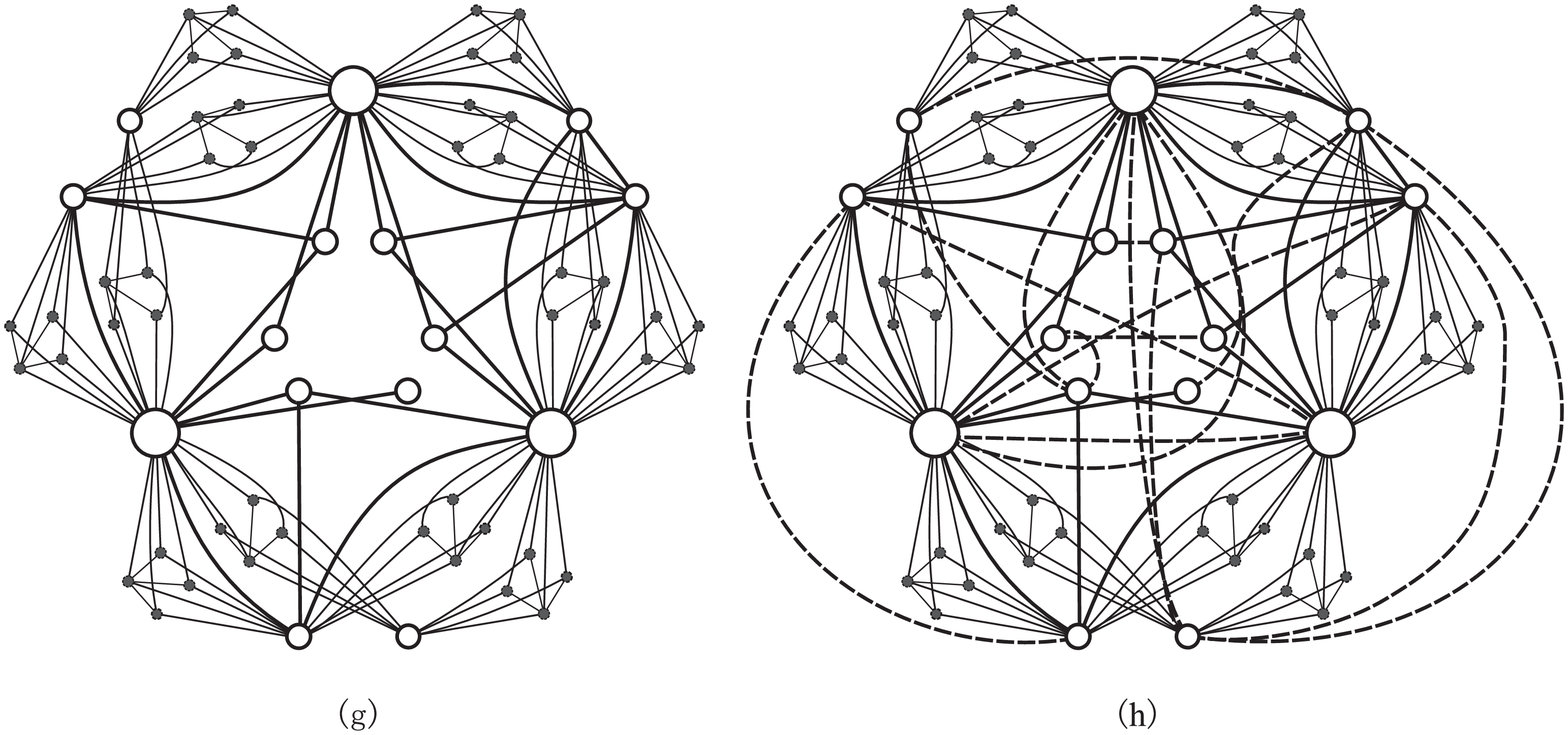}
\caption{\label{fig:randomized-example-3} {\small (g) $T'(1)$ is obtained by removing $12$ bold-edges of $N'(1)$ shown in Fig.\ref{fig:randomized-example-1} shown in Fig.\ref{fig:randomized-example-1} by the probability $p_r=\frac{1}{3}$; (h) the randomized EBGN-model $N'(2)$ obtained by adding $16$ new dashing-edges to $T'(1)$ by the probability $p_a=\frac{2}{3}$. In this case, the number of edges of $N'(2)$ is greater than that of $N'(1)$.}}
\end{figure}

Notice that $N'(t-1)$ is not a subgraph of $N'(t)$ by the construction of the Algorithm-I. We call the Algorithm-I the \emph{growth-first randomness-second algorithm}. An illustration about the model $N'(t)$ is shown in Fig.\ref{fig:randomized-example-1}, Fig.\ref{fig:randomized-example-2} and Fig.\ref{fig:randomized-example-3}. We have the following basic parameters
\begin{equation}\label{eqa:II-three-basic-numbers}
{
\begin{split}
N'_{v,t}&=N'_{v,0}+m_v N'_{e,0} \sum ^{t-1}_{k=0}(2r)^k=N'_{v,0}+ m_v N'_{e,0}\frac{(2r)^{t}-1}{2r-1},\\
N'_{e,t}&=p^{t-1}+(m_e+2m_v)N'_{e,0}\frac{(2r)^{t}-p^{t}}{2r-p},\\
N'_{be,t}&=(2r)^t N'_{e,0}.
\end{split}}
\end{equation}
where $p=(1-p_r)(1+p_a)$. Here, two numbers $N'_{v,t}$ and $N'_{be,t}$ are shown in (\ref{eqa:three-basic-numbers}), and the edge number $N'_{e,t}$ will be deduced in the proof of Theorem \ref{them:main-thereom} in Appendix.

Let $\langle k\rangle '_0=\frac{2m_e}{m_v}$ be the \emph{average degree} of the initial model $N'(0)$. By Eq.(\ref{eqa:II-three-basic-numbers}), the average degree $\langle k\rangle'$ of $N'(t)$ is easily estimated as follows.
\begin{equation}\label{eqa:new-average-degree}
\langle k\rangle'=\frac{2N'_{e,t}}{N'_{v,t}}\propto \frac{(m_e+2m_v)(2r-1)}{m_v(2r-p)}=\left (\frac{\langle k\rangle '_0}{2}+2\right )\frac{2r-1}{2r-p},\qquad t\rightarrow \infty.
\end{equation}
The result (\ref{eqa:new-average-degree}) shows that $N'(t)$ is a tree-like model.
Since $p_r=\frac{\textrm{the number of removing edges}}{N'_{e,t-1}}$ and the number of adding edges is equal to $p_a(1-p_r)N'_{e,t-1}$, we announce that $p_a\geq \frac{p_r}{1-p_r}$ as $0<p_a\leq \frac{1}{2}$, and $0<p_a<\frac{p_r}{1-p_r}$ when $\frac{1}{2}<p_r<1$ for the existence of the randomized EBGN-model $N'(t)$. We define the \emph{edge-cumulative distribution} $P'_{\text{e-cum}}(k)$ of the model $N'(t)$ by $$P'_{\text{e-cum}}(k)=\frac{1}{N'_{e,t}}\left (\sum^{\tau}_{i=0} N'_{e,i}\right )$$ with $0<\tau<t$. We can show a result in Theorem \ref{them:main-thereom} whose proof is detailed in Appendix:
\begin{thm} \label{them:main-thereom}
The randomized EBGN-model $N'(t)$ is scale-free, and its edge-cumulative distribution $P'_{\text{e-cum}}(k)$ obeys the power law distribution and is independent of $p=(1-p_r)(1+p_a)$.
\end{thm}

\section{Models having no edges removed/added}

We construct our $(r,F)$-ugnms $N(t)$ for time steps $t\geq 0$ in the following deterministic algorithm-II. The seed set $F$ is defined well and the bound thickness $r$ subject to $m_v\geq r\geq 1$ in Section 2.
\begin{verse}
\textbf{Algorithm-II}

\textbf{Step 1.} (Initialization) For $t=0$, $N(0)$ is a connected network model having no multiple edges and loops. $N(0)$ has its own vertex-set $V(0)$ with $N_{v,0}=|V(0)|~(\geq 2)$ vertices and edge-set $E(0)$ with $N_{e,0}=|E(0)|~(\geq 1)$ edges. We define every edge of $N(0)$ as a \emph{bound-edge} and call $N(0)$ the \emph{initial $(r,F)$-ugnm}. Let $B(0)$ be the bound-edge set of $N(0)$.

\textbf{Step 1.} (Iteration) For $t\geq 1$, an $(r,F)$-ugnm $N(t)$ is obtained by doing the following operation-II:

\textbf{Operation-II:} Add a \emph{seed} $G\in F$ to each bound-edge $uv$ of $N(t-1)$, and joining every vertex $x_{i}$ of $G$ with vertex $u$ and vertex $v$, respectively, produces two edges $x_{i}u$ and $x_{i}v$; and select arbitrarily vertices $x_{i_1},x_{i_2},\dots, x_{i_r}$ to define $2r$ edges $x_{i_j}u$ and $vx_{i_j}$ for $j\in [1, r]$ as the bound-edges of $N(t)$.
\end{verse}

Thereby, we write $X(t)=V(t)\setminus V(t-1)$ to indicate the set of new vertices added into $N(t-1)$, $Y(t)=E(t)\setminus E(t-1)$ the set of new edges added into $N(t-1)$, and $B(t)$ the set of bound-edges of $N(t)$. Clearly, different bound-edges of $N(t-1)$ may correspond to different seeds in $F$. It is not difficult to observe that $N(t-1)$ is a sub-model of $N(t)$ for $t\geq 1$, so we also call $N(t)$ a \emph{nested $(r,F)$-ugnm}.

\subsection{Basic properties}

We denote the degree of a vertex $x$ of an $(r,F)$-ugnm $N(t)$ at time step $t$ as $\ud (x,t)$ that is the number of edges incident to $x$ in $N(t)$, and write the numbers of vertices, edges and bound-edges of $N(t)$ by $N_{v,t}=|V(t)|$, $N_{e,t}=|E(t)|$ and $N_{be,t}=|B(t)|$, respectively.

\textbf{Basic numbers.} By the construction of an $(r,F)$-ugnm $N(t)$, we are not difficult to obtain
\begin{equation}\label{eqa:initial-formula}
N_{v,1}=N_{v,0}+m_v N_{e,0} ,~N_{e,1}=N_{e,0}+ (m_e+2m_v)N_{e,0},~N_{be,1}=2r N_{be,0},
\end{equation}
where $N_{be,0}=N_{e,0}$. For $k\geq 2$, we have the following recursive formulas
\begin{equation}\label{eqa:recursive-formulas}
{
\begin{split}
N_{v,k}&=N_{v,k-1}+m_v N_{be,k-1} ,\\
N_{e,k}&=N_{e,k-1}+(m_e+2m_v)N_{e,k-1},\\
N_{be,k}&=2r N_{be,k-1}.
\end{split}}
\end{equation}
Furthermore, the numbers $N_{v,t},N_{e,t},N_{be,t}$ of vertices, edges and bound-edges of an $(r,F)$-ugnm $N(t)$ at time step $t\geq 1$ are as follows:
\begin{equation}\label{eqa:three-basic-numbers}
{
\begin{split}
N_{v,t}&=N_{v,0}+ m_v N_{e,0}\frac{(2r)^{t}-1}{2r-1},\\
N_{e,t}&=N_{e,0}+(m_e+2m_v)N_{e,0}\frac{(2r)^{t}-1}{2r-1},\\
N_{be,t}&=(2r)^t N_{e,0}.
\end{split}}
\end{equation}
Eq.(\ref{eqa:three-basic-numbers}) shows $N(t)$ has the growth behavior as $t$ tends to infinity.

\textbf{Average degree.} The \emph{average degree} $\langle k\rangle$ of an $(r,F)$-ugnm $N(t)$ at time step $t\geq 1$ can be estimated as
\begin{equation}\label{eqa:average-degree}
{
\begin{split}
\langle k\rangle&=\frac{2N_{e,t}}{N_{v,t}}=2\cdot \frac{N_{e,0}+(m_e+2m_v)N_{e,0}\sum ^{t-1}_{k=0}(2r)^k}{N_{v,0}+ m_v N_{e,0}\sum ^{t-1}_{k=0}(2r)^k}\\
&\propto \frac{2(m_e+2m_v)}{m_v}=4+\langle k\rangle_0,
\end{split}}
\end{equation}
for larger $t$, where $\langle k\rangle_0=\frac{2m_e}{m_v}$ is the average degree of the initial $(r,F)$-ugnm $N(0)$. Clearly, an $(r,F)$-ugnm $N(t)$ is a \emph{sparse model} for larger $t$, since $\langle k\rangle_0$ is a constant. Also, $N(t)$ is tree-like, since the number $N_{e,t}$ of edges of $N(t)$ grows linearly with the number $N_{v,t}$ of vertices of $N(t)$ at time step $t$.

Notice that two numbers of the newly added vertices and the newly added edges in forming an $(r,F)$-ugnm $N(t)$ are
\begin{equation}\label{eqa:number-new-vertex-edge}
{
\begin{split}
&|X(t)|=|V(t)\setminus V(t-1)|=m_v (2r)^{t-1}N_{e,0},\\
&|Y(t)|=|E(t)\setminus E(t-1)|=2|X(t)|=(m_e+2m_v)(2r)^{t-1}N_{e,0},
\end{split}}
\end{equation} respectively. The new increasing part of an $(r,F)$-ugnm $N(t)$ at time step $t$ is tree-like, that is, the number $|Y(t)|$ of new edges added to $N(t-1)$ grows linearly with the number $|X(t)|$ of vertices newly added according to $\frac{|Y(t)|}{|X(t)|}\propto 2$ as $t$ tends to infinity.

\subsection{Degree spectrum, diameter}
By the \emph{linear preferential attachment rule} in Ref.\cite{Costa-Rodrigues-Travieso-Villas-Boas2006}, the vertices having larger degrees in $N(t)$ play the role of connecting network and attract easily new vertices that are entering into $N(t)$. A vertex $w$ added newly is easily adjacent to those vertices $z$ of the network $N(t)$ having larger degrees, that is,
\begin{equation}\label{eqa:linear-preferential-attachment-rule}
P_t(w\rightarrow z)=\frac{\ud (z,t)}{\sum_{v\in V(t)}\ud (v,t)}=\frac{\ud (z,t)}{2N_{e,t}}.
\end{equation}
So, finding the degrees of vertices is very important for figuring out many properties of networks.

For a vertex $u$ in the initial model $N(0)$, it is not hard to compute its degree in $N(t)$
\begin{equation}\label{eqa:degree-formula}
\ud (u,t)=\left\{
\begin{array}{ll}
\displaystyle \left (1+m_v\frac{r^t-1}{r-1}\right )\ud (u,0),&r\geq 2;\\[8pt]
(1+tm_v)\ud (u,0),&r=1.
\end{array}
\right.
\end{equation}And the vertex $u$ is the common end of $r^t\ud (u,0)$ bound-edges of $N(t)$ when $t\geq 1$.

In forming $N(s)$ at time step $s\geq 1$ with respect to $s\leq t$, a vertex $x$ of a seed $G\in F$ is added newly into $N(s-1)$ and is selected as an end of some bound-edge of $N(s)$. So, in $N(t)$ at time step $t\geq s\geq 1$, the vertex $x$ has its own degree
\begin{equation}\label{eqa:degree-formula11}
\ud (x,t)=\left\{
\begin{array}{ll}
\displaystyle \ud (x,G)+2+2m_v\frac{r^{t-s}-1}{r-1},&r\geq 2;\\[8pt]
\ud (x,G)+2+2m_v(t-s),&r=1,
\end{array}
\right.
\end{equation} and is the common end of $2r^{t-s}$ bound-edges of $N(t)$.

For a vertex $y$ of a seed $G\in F$ being not an end of any bound-edge of $N(t)$, we know its own degree $\ud (y,t)=\ud (y,G)+2$. Eq.s (\ref{eqa:linear-preferential-attachment-rule}), (\ref{eqa:degree-formula}) and (\ref{eqa:degree-formula11}) show that our $(r,F)$-ugnm $N(t)$ possess the behavior of preferential attachment, since for larger $t$ and $r\geq 2$, we have $$P_t(w\rightarrow u)>P_t(w\rightarrow x)>P_t(w\rightarrow y).$$

We can rewrite new added-vertex set $X(s)$ at each time step $s\in [1,t-1]$ with $t\geq 2$ by $X(s)=X^{be}_s\cup X^{nbe}_s$, where $X^{be}_s$ is the set of newly added vertices that are selected as the ends of some bound-edges of $N(s)$, and every vertex of the set $X^{nbe}_s=X(s)\setminus X^{be}_s$ is not an end of any bound-edge. It is not difficult to compute both numbers $|X^{be}_s|$ and $|X^{nbe}_s|$ by Eq.(\ref{eqa:number-new-vertex-edge}). We can see that $|X^{be}_s|=rN_{e,0}(2r)^{s-1}$, and the degree of each $y_i$ of $X^{be}_s$ in $N(t)$ is shown in Eq.(\ref{eqa:degree-formula11}); $|X^{nbe}_s|=(m_v-r)N_{e,0}(2r)^{s-1}$, each $y_j$ of $X^{nbe}_s$ has its own degree $\ud (y_j,G)+2$ in $N(t)$.

By the above degree spectrum of $N(t)$ we can compute the probability $P(k)$  that a randomly selected node has
exactly $k$ edges. Suppose that each vertex of $X^{be}_s$ has its own degrees greater than $k$ for $s\in [1,\tau]$ with $0<\tau <t$, and $P(k^*>k)$ is the probability of vertices having degree not less than $k$. So, by (\ref{eqa:degree-formula}) and (\ref{eqa:degree-formula11}), we have $\ud (u,t)>k$ for $u\in V(0)$, and $\ud (x,s)>k$ for $x\in X^{be}_s$ with $s\in [1,\tau]$, but $X^{be}_s$ contains no vertex $x$ having its own degree $\ud (x,s)>k$. Thereby,
\begin{equation}\label{eqa:probability-P(k)}
{
\begin{split}
P(k^*>k)&=\frac{1}{N_{v,t}}\left [N_{v,0}+\sum^{\tau}_{s=1}|X^{be}_s|\right ]=\frac{1}{N_{v,t}}\left [N_{v,0}+rN_{e,0}\sum^{\tau}_{s=1}(2r)^{s-1}\right ]\\
&=\left [N_{v,0}+rN_{e,0}\frac{(2r)^{\tau }-1}{2r-1}\right ]\Big /\left [N_{v,0}+N_{e,0} m_v\frac{(2r)^{t}-1}{2r-1}\right ]\\
&\propto \frac{r}{m_v}(2r)^{\tau-t},
\end{split}}
\end{equation}
and $P(k^*>k-1)\propto \frac{r}{m_v}(2r)^{\tau+1-t}$, which enable us to get
\begin{equation}\label{eqa:probability-1}
P(k)=P(k^*>k-1)-P(k^*>k)\propto \frac{r(2r-1)}{m_v}(2r)^{\tau-t}.
\end{equation}
Let $h(r)=\frac{2r}{\ln (2r)}+\frac{1}{\ln (2r+m_v)}$. Plugging $\tau=t-h(r)\ln k$ into Eq.(\ref{eqa:probability-1}) can show
$$P(k)\propto \frac{r(2r-1)}{m_v}k^{-h(r)\ln (2r)},$$
since
\begin{equation}\label{eqa:c3-scale-free-number}
(2r)^{\tau-t}=(2r)^{-h(r)\ln k}
=k^{-h(r)\ln (2r)}.
\end{equation} Hence, $N(t)$ is an exponentially growing model, also, is a scale-free model as mentioned in \cite{Albert-Barabasi-family2002}.

\textbf{Diameter.} The notation $\text{dis}(x,y)$ indicates the distance between two vertices $x$ and $y$ of $N(t)$, and $D(t)$ denotes the diameter of $N(t)$ at time step $t$. Note that $\text{dis}(x,y)$ is the length of a shortest path connecting $x$ and $y$, and $D(t)=\max\{\text{dis}(x,y):~x,y\in V(t)\}$. For a new added vertex $x\in X(s)$ at time step $s\geq 2$, the algorithmic construction of $N(t)$ tells us that $x$ is adjacent to two vertices $u_i$ and $u_j$, where $u_i\in X(i)$ and $u_j\in X(j)$ with $i\neq j$ such that $(i,j)\in \{(0,s-1),(1,s-1),(2,s-1),\dots, (s-2,s-1)\}$, and $X(0)=V(0)$. So, $x$ can be connected with a vertex $u_0$ of the initial model $N(0)$ by a path $xu_{s-2}u_{s-4}\cdots u_2u_0$ or another path $xu_{s-1}u_{s-3}\cdots u_1u_0$. Thereby, two vertices $x'$ and $x''$ of $N(t)$ can be connected with two vertices $u_0$ and $v_0$ of $N(0)$ by two paths $P(x',u_0)=x'u_{i-2}u_{i-4}\cdots u_2u_0$ and $Q(x'',v_0)=x''u_{j-2}u_{j-4}\cdots u_2v_0$, respectively. Let $U(u_0,v_0)$ be the shortest path connecting $u_0$ and $v_0$ in $N(0)$, so its length $|U(u_0,v_0)|\leq D(0)$. We can estimate the distance $\text{dis}(x',x'')$ in the following inequalities
$$\text{dis}(x',x'')\leq |P(x',u_0)|+|U(u_0,v_0)|+|Q(x'',v_0)|\leq 2\left \lfloor \frac{t+1}{2}\right \rfloor +D(0),$$
this result shows $D(t)\leq t+1+D(0)$ at time step $t$. Therefore, $N(t)$ is small-world because its diameter $D(t)$ is the same rank as $\log N_{v,t}$, that is, $D(t)=O(\log N_{v,t})$.

\subsection{New properties}

We present some new statistical approaches for exploring connections between known and new statistics methods.

\textbf{Newly added average degree.} We can get the \emph{newly added average degree} $\langle k\rangle_{new}$ as follows.
\begin{equation}\label{eqa:newly-added-average-degree}
\langle k\rangle_{new}=\frac{2|Y(t)|}{|X(t)|}=\frac{2|E(t)\setminus E(t-1)|}{|V(t)\setminus V(t-1)|}=\frac{2(m_e+2m_v)(2r)^{t-1}N_{e,0}}{ m_v (2r)^{t-1}N_{e,0}}=4+\langle k\rangle_0\propto \langle k\rangle.
\end{equation}
Clearly, two averages $\langle k\rangle_{new}$ and $\langle k\rangle$ are the part and the whole, respectively; so that the part $\langle k\rangle_{new}$ is equivalent to the whole $\langle k\rangle$ according to Eq.(\ref{eqa:average-degree}) and Eq.(\ref{eqa:newly-added-average-degree}) for sufficiently large $t$.

\textbf{Edge-cumulative distribution.} For an integer $\delta$ with respect to $0< \delta <t$, we define the \emph{edge-cumulative distribution} $P_{\textrm{e-cum}}(k)$ of an $(r,F)$-ugnm $N(t)$ by $P_{\textrm{e-cum}}(k)=\frac{1}{N_{e,t}}\sum ^{\delta}_{s=0}N_{e,s}$. By Eq.(\ref{eqa:three-basic-numbers}) we have the following result.
\begin{thm} \label{them:theorem-112}
The edge-cumulative distribution $P_\textrm{{e-cum}}(k)$ of the $(r,F)$-ugnm $N(t)$ obeys the power law distribution, since
\begin{equation}\label{eqa:edge-cumulative-theorem}
P_\textrm{{e-cum}}(k)\propto\frac{2r}{2r-1} k^{-h(r)\ln (2r)}.
\end{equation}
\end{thm}
The proof of Theorem \ref{them:theorem-112} is developed in Appendix.

When $N(0)$'s vertex number $m_v\geq 1$ and the bound thickness $r=1$ we obtain $\gamma_k=2+\frac{\ln 2}{\ln (2+m_v)}$ with $2<\gamma_k<3$. Clearly, the case $r\geq 2$ shows $h(r)\ln (2r)=2r+\frac{\ln (2r)}{\ln (2r+m_v)}>4$. In other words, the edge-cumulative distribution $P_\textrm{{e-cum}}(k)$ will be influenced only by the parameter $r$. We claim that there is a function $f(r,m_v)$ such that $P(k)=f(r,m_v) P_\textrm{{e-cum}}(k)$, according to the following
\begin{equation}\label{eqa:c3-key-form}
\frac{P(k)}{P_\textrm{{e-cum}}(k)}\propto \frac{(2r-1)^2}{2m_v},\quad \textrm{or}\quad P(k)\propto \frac{(2r-1)^2}{2m_v}P_\textrm{{e-cum}}(k).
\end{equation}

Observe that $P(k^*\leq k)$ and $P_\textrm{{e-cum}}(k)$, where $P(k^*\leq k)$ is the probability of vertices having degrees less than $k+1$. By $P(k^*\leq k)=1-P(k^*> k)\propto 1-\frac{r}{m_v}(2r)^{\tau -t}$, we have \begin{equation}\label{eqa:c3xxxxx}
\frac{P(k^*\leq k)}{P_\textrm{{e-cum}}(k)}\propto \frac{(m_v-r)(2r-1)}{2rm_v},~t\rightarrow \infty.
\end{equation}

\textbf{The $(v_k,e_k)$-models.} Based on the degree spectrum of an $(r,F)$-ugnm $N(t)$, we will compute the number $S_N(\leq k)$ of vertices having degrees no more than $k$ in an $(r,F)$-ugnm $N(t)$ and the sum $Q_N(\leq k)$ of degrees of these vertices at time step $t$. Clearly, the number of vertices having degrees greater than $k$ is equal to $S_N(> k)=N_{v,t}-S_N(\leq k)$ and the number of degrees of the vertices whose degrees are greater than $k$ is equal to $Q_N(>k)=2N_{e,t}-Q_N(\leq k)$. Here, we consider a particular selection as forming $N(t)$. Suppose that every graph $G\in F$ has the vertices $y_1,y_2,\dots ,y_{m_v}$ and a vertex $y_i$ has its own degree $\ud (y_i,G)=d_i$ for $i\in [1,m_v]$ such that $d_j\geq d_{j+1}$ for $i\in [j,m_v-1]$. In the procedure of constructing an $(r,F)$-ugnm $N(t)$ by adding a \emph{seed} $G$ to each bound-edge $uv$ of $N(t-1)$, we select the previous vertices $y_{1},y_{2}, \dots ,y_{r}$ of the seed $G$ and define $2r$ edges $y_{j}u$ and $vy_{j}$ as the bound-edges of $N(t)$ for $j\in [1,r]$.

\begin{thm} \label{them:theorem-222}
For $r\geq 2$, two quantities $v_k=\frac{S_N(> k)}{N_{v,t}}$ and $e_k=\frac{Q_N(>k)}{2N_{e,t}}$ obey the power law distribution according to
$${
\begin{split}
v_k&\propto \frac{r}{m_v}k^{-h(r)\ln (2r)},\\
e_k&\propto \left [\frac{m^*_e+2r}{m_e+2m_v}-\frac{rm_v}{(r-1)(2r-1)}\right ]\cdot \frac{k^{-h(r)\ln (2r)}}{2}+\frac{ m_v(2r-1)}{m_e+2m_v}\cdot \frac{k^{-h(r)\ln 2}}{r-1},
\end{split}}
$$ where $h(r)=\frac{2r}{\ln (2r)}+\frac{1}{\ln (2r+m_v)}$ and $m^*_e=\sum^r_{i=1}\ud (y_i,G)$.
\end{thm}
The proof of Theorem \ref{them:theorem-222} is presented in Appendix.

\section{Derivative models}

Our uniformly $(r,F)$-growing network model $N(t)$ can be used to build up another class of randomized model $M'(t)$ having partially rewiring edges in the way that we do an operation-II to each bound-edge of $M'(t-1)$ and next, rewire some edges of $M'(t-1)$. Notice that the operation-II guarantees the growth and preferential attachment of both models $N(t)$ and $M'(t)$, and two models both have the same number of edges. By means of the edge-cumulative distribution, we can demonstrate that $M'(t)$ obeys the power-law distribution.

One more randomized bound-growing network model $M''(t)$ can be defined by applying the step 1 of the Algorithm-I, and then, for $t\geq 2$, $M''(t)$ is obtained by doing an operation-I to each bound-edge of $M''(t-1)$ and make the number $M''_{e,t}$ of edges of $M''(t)$ holds $N'_{e,t}\leq M''_{e,t}\leq N_{e,t}$ (or $N_{e,t}\leq M''_{e,t}\leq N'_{e,t}$). So, the scale-free behavior of $M''(t)$ can be proven through comparing three edge-cumulative distributions of $M''(t)$, $N'(t)$ and $N(t)$.


\section{Conclusion}

We, by means of the topological structure of our uniformly $(r,F)$-growing network model $N(t)$, show the scale-free behavior of our randomized EBGN-model $N'(t)$. However, it is not easy to obtain the exact values of some distributions of the randomized EBGN-model $N'(t)$, such as the power-law degree distribution, the clustering coefficient distribution and diameter, since we are not able to calculate its degree spectrum and $N'(t)$ may be disconnected. To prove the randomized EBGN-model $N'(t)$ to be scale-free, we use the principle of comparison to show that two edge-cumulative distributions $P_{\text{e-cum}}(k)$ and $P'_{\text{e-cum}}(k)$ both are equivalent to each other. In contrast, the removing-edge probability $p_r$ and the adding-edge probability $p_a$ hold $p_r<\frac{1}{2}<p_a<1$, the number of edges of the randomized EBGN-model $N'(t)$ is greater than the number of edges of the uniformly $(r,F)$-growing network model $N(t)$, that is, $N'_{e,t}>N_{e,t}$; and moreover $N'_{e,t}<N_{e,t}$ when $p_a<\frac{1}{2}<p_r<1$. Likewise, we observe that $N'(t)$ has a giant component that charges the growth and preferential attachment and makes $N'(t)$ to be scale-free; the other part of $N'(t)$, called the subpart, does not determine the topological structure of $N'(t)$ no matter removing edges from the subpart or adding edges to the subpart. We use a notation $G(t-1)$ to denote the remainder after removing and adding edges to $N'(t-1)$ in the Algorithm-I. It may occur a phenomenon that $G(t-1)$ is regular rather than scale-free at some time step $t\geq 1$, that is, $G(t-1)$ is a random Erd\"{o}s-Reny\'{i} type graph that follows the form of the Poisson variable.

We point out the fragility of our randomized EBGN-model $N'(t)$. If we remove the vertices of the initial model $N'(0)$ from $N'(t)$, the remainder may have two or more components that are connected to each other. For example, attacking the vertices of the initial model $N'(0)$ at time step $t_0$ may make the remainder $M(t_0)$ of $N'(t_0)$ has some components $M_i(t_0;t)$ for $i=1,2,\dots ,m~(\geq 2)$. However, according to our construction algorithm of $N'(t)$, a component $M_i(t_0;t)$ having some bound-edges can be self-growing and obeys the power-law distribution for enough larger $t>t_0$, so that the model $M(t_0;t)$ based on the initial model $M(t_0)$ will become a scale-free model as adding edges make the model $M(t_0;t)$ to be connected.

Clearly, the models mentioned above can be generalized into directed models. As further work we propose:

(1) If, to obtain $N'(t)$, we do firstly removing/adding edges to $N'(t-1)$, and then grow the remainder by doing the operation-I to each bound-edge of the remainder. This construction algorithm is refereed as \emph{randomness-first growth-second}. Is the model $N'(t)$ scale-free?

(2) In the step 2 of the Algorithm-I, we substitute by $p_{r,t}$ and $p_{a,t}$ two probabilities $p_r$ and $p_a$ at time step $t$, such that there are $p_{r,i}\neq p_{r,j}$ or $p_{a,s}\neq p_{a,l}$ for some $i\neq j$ or $s\neq l$. What topological structure does $N'(t)$ have? In other words, we want to optimize randomized EBGN-model $N'(t)$, as proposed in \cite{David-Alderson2008}.

\vskip 0.4cm

\textbf{Acknowledgment.}
This work was funded partly by the National Natural Science Foundation of China under No. 61163054, No. 61363060 and No. 61163037.


\vskip 1cm

{\Large \textbf{Appendix}}

\vskip 0.4cm

\textbf{Proof of Theorem \ref{them:theorem-112}.} To calculate the edge-cumulative distribution of $N(t)$, we take $\delta$ with respect to $0<\delta <t$ for larger $t$. So
\begin{equation}\label{eqa:edge-cumulative-distribution}
{
\begin{split}
P_{\textrm{e-cum}}(k)&=\frac{1}{N_{e,t}}\sum ^{\delta}_{s=0}n_{e,s}\\
&=\frac{1}{N_{e,t}}\left \{N_{e,0}+\sum ^{\delta}_{s=1}\left [N_{e,0}+N_{e,0}(m_e+2m_v)\sum ^{s-1}_{j=0}(2r)^j\right ]\right \}\\
&=\frac{(1+\delta )N_{e,0}}{N_{e,t}}+\frac{(m_e+2m_v)N_{e,0}}{N_{e,t}}\sum ^{\delta}_{s=1}\sum ^{s-1}_{j=0}(2r)^j\\
&=\frac{(1+\delta )N_{e,0}}{N_{e,t}}+\frac{(m_e+2m_v)N_{e,0}}{N_{e,t}}\sum ^{\delta}_{s=1}\frac{(2r)^s-1}{2r-1}\\
&=\frac{(1+\delta )N_{e,0}}{N_{e,t}}+\frac{(m_e+2m_v)N_{e,0}}{(2r-1)N_{e,t}}\left [\frac{(2r)^{1+\delta }-1}{2r-1}-(1+\delta )\right ]\\
&\propto \frac{(2r)^{1+\delta -t}}{2r-1}=\frac{2r}{2r-1}(2r)^{\delta -t}
\end{split}}
\end{equation}
Plugging $\delta=t-h(r)\ln k$ for $h(r)=\frac{2r}{\ln (2r)}+\frac{1}{\ln (2r+m_v)}$ into Eq.(\ref{eqa:edge-cumulative-distribution}) yields
\begin{equation}\label{eqa:N(t)-edge-cumulative}
P_\textrm{{e-cum}}(k)\propto\frac{2r}{2r-1} k^{-h(r)\ln (2r)}
\end{equation}
by Eq.(\ref{eqa:c3-scale-free-number}), as desired.\js

\vskip 0.4cm

\textbf{Proof of Theorem \ref{them:theorem-222}.} For $r\geq 2$, we can estimate
$$\frac{S_N(\leq k)}{N_{v,t}}=P(k^*\leq k)=1-P(k^*> k)=1-\frac{r}{m_v}(2r)^{\tau -t}$$
by the form (\ref{eqa:probability-P(k)}) and Eq.(\ref{eqa:c3-scale-free-number}). Because of $\frac{S_N(>k)}{N_{v,t}}=1-\frac{S_N(\leq k)}{N_{v,t}}$, so we have \begin{equation}\label{eqa:v-k-variable}
v_k=\frac{S_N(>k)}{N_{v,t}}\propto \frac{r}{m_v}k^{-h(r)\ln (2r)}.
\end{equation}

As $r\geq 2$, we come to compute
\begin{equation}\label{eqa:more-deduction}
Q_N(>k)=\sum^{N_{v,0}}_{i=1}\ud (u_i,t)+\sum^{\tau}_{s=1}\sum_{y_i\in X^{be}_s}\ud (y_i,t)
\end{equation}
The first term in Eq.(\ref{eqa:more-deduction}) can be compute as
$$
\sum^{N_{v,0}}_{i=1}\ud (u_i,t)=\sum^{N_{v,0}}_{i=1}\left (1+m_v\frac{r^t-1}{r-1}\right )\ud (u,G)=2N_{e,0}\left (1+m_v\frac{r^t-1}{r-1}\right );$$
the second term of Eq.(\ref{eqa:more-deduction}) can be calculated as
$$
{
\begin{split}
&\quad \sum^{\tau}_{s=1}\sum_{y_i\in X^{be}_s}\ud (y_i,t)=\sum^{\tau}_{s=1}\sum^r_{i=1}N_{e,0}(2r)^{s-1}\left [\ud (u,G)+2+m_v\frac{r^{t-s}-1}{r-1}\right ]\\
&=N_{e,0}\left (\sum^{\tau}_{s=1}\sum^r_{i=1}(2r)^{s-1}\ud (y_i,G)+2r\sum^{\tau}_{s=1}(2r)^{s-1}+\frac{rm_v}{r-1}\sum^{\tau}_{s=1}(2r)^{s-1}[r^{t-s}-1]\right )\\
&=N_{e,0}\left (\sum^r_{i=1}\sum^{\tau}_{s=1}(2r)^{s-1}\ud (y_i,G)+2r\frac{(2r)^{\tau}-1}{2r-1}+\frac{rm_v}{r-1}\left [(2^{\tau}-1)r^{t-1}-\frac{(2r)^{\tau}-1}{2r-1}\right ]\right )\\
&=N_{e,0}\left (\frac{(2r)^{\tau}-1}{2r-1}\sum^r_{i=1}\ud (y_i,G)+2r\frac{(2r)^{\tau}-1}{2r-1}+\frac{rm_v}{r-1}\left [(2^{\tau}-1)r^{t-1}-\frac{(2r)^{\tau}-1}{2r-1}\right ]\right )
\end{split}}
$$
Let $m^*_e=\sum^r_{i=1}\ud (y_i,G)$. We can estimate
\begin{equation}\label{eqa:new-models}
{
\begin{split}
e_k=\frac{Q_N(>k)}{2N_{e,t}}&\propto \frac{\sum^{N_{v,0}}_{i=1}\ud (u_i,t)+\frac{(2r)^{\tau}-1}{2r-1}m^*_e+2r\frac{(2r)^{\tau}-1}{2r-1}+\frac{rm_v}{r-1}\left [(2^{\tau}-1)r^{t-1}-\frac{(2r)^{\tau}-1}{2r-1}\right ]}{2(m_e+2m_v)\frac{(2r)^{t}-1}{2r-1}}\\
&\propto \frac{m_v}{m_e+2m_v}\frac{1}{2^t}+\left [\frac{m^*_e+2r}{m_e+2m_v}-\frac{rm_v}{(r-1)(2r-1)}\right ]\cdot \frac{(2r)^{\tau-t}}{2}\\
&\quad +\frac{r m_v(2r-1)}{(m_e+2m_v)(r-1)}\cdot \frac{2^{\tau-t}}{2r}.
\end{split}}
\end{equation}
Notice that $\frac{m_v}{2^t(m_e+2m_v)}\rightarrow 0$ as $t\rightarrow \infty$. Plugging $\tau=t-h(r)\ln k$ into Eq.(\ref{eqa:new-models}) produces
\begin{equation}\label{eqa:a-k-variable}
e_k=\frac{Q_N(>k)}{2N_{e,t}}\propto A\cdot k^{-h(r)\ln (2r)}+B\cdot k^{-h(r)\ln 2}.
\end{equation}
where $A=\frac{1}{2}\left [\frac{m^*_e+2r}{m_e+2m_v}-\frac{rm_v}{(r-1)(2r-1)}\right ]$ and $B=\frac{m_v(2r-1)}{(r-1)(m_e+2m_v)}$. Note that $m^*_e\leq 2m_e$. The proof of this theorem is complete.\js

\vskip 0.4cm

\textbf{The proof of Theorem \ref{them:main-thereom}.} Clearly,
$N'_{e,1}=(1+m_e+2m_v)N'_{e,0}$, and we have known the value of $N_{e,k}$ of an $(r,F)$-ugnm $N'(t)$ by Eq.(\ref{eqa:three-basic-numbers}). For $s\geq 2$, the edge number $N'_{e,s}$ of $N'(s)$ has three parts: the number of newly added edges of $Y(s)$, the number $N'_{e,s-1}-p_{r}N'_{e,s-1}=(1-p_{r})N'_{e,s-1}$ of randomly removed edges of $N'(s-1)$ with probability $p_{r}$ and the number $p_{a}(1-p_{r})N'_{e,s-1}$ of new edges randomly added to the remainder obtained from $N'(s-1)$ by randomly removing edges of $N'(s-1)$ according to probability $p_{a}$. Let $p=(1-p_{r})(1+p_{a})$. So, we have a recursive formula
\begin{equation}\label{eqa:edge-number-randomized-model}
N'_{e,s}=|Y(s)|+pN'_{e,s-1}.
\end{equation}
Note that $N'(t)$ and $N(t)$ have the same number of new vertices added and the same number of new edges added to $N'(t-1)$ and $N(t-1)$, respectively. By Eq.(\ref{eqa:number-new-vertex-edge}) we use repeatedly the recursive formula (\ref{eqa:edge-number-randomized-model}), finally, we obtain
\begin{equation}\label{eqa:randomly-models}
{
\begin{split}
&\quad N'_{e,t}=p^{t-1}N'_{e,1}+\sum^{t-2}_{s=0}p^{s}|Y(t-s)|\\
&=p^{t-1}(1+m_e+2m_v)N'_{e,0}+(m_e+2m_v)N'_{e,0}\sum^{t-2}_{s=0}p^{s}(2r)^{t-s-1}\\
&=p^{t-1}+(m_e+2m_v)N'_{e,0}\frac{(2r)^{t}-p^{t}}{2r-p}.
\end{split}}
\end{equation}
We compute the edge-cumulative distribution of $N'(t)$ for $0<\tau <t$ as follows:
$$
{
\begin{split}
&\quad P'_{\text{e-cum}}(k)=\frac{1}{N'_{e,t}}\left (N'_{e,0}+N'_{e,1}+\sum ^{\tau}_{s=2}N'_{e,s}\right )\\
&=\frac{1}{N'_{e,t}}\left \{(2+m_e+2m_v)N'_{e,0}+\sum ^{\tau}_{s=2}p^{s-1}+\frac{(m_e+2m_v)N'_{e,0}}{2r-p}\sum ^{\tau}_{s=2}[(2r)^{s}-p^{s}]\right \}\\
&=\frac{1}{N'_{e,t}}\left \{(2+m_e+2m_v)N'_{e,0}+\frac{p^{\tau}-p}{p-1}+\frac{(m_e+2m_v)N'_{e,0}}{2r-p}\left [\frac{(2r)^{\tau }-2r}{2r-1}-\frac{p^{\tau}-p}{p-1}\right ]\right \}
\end{split}}
$$
Since $\left (\frac{p}{2r}\right )^{t}\rightarrow 0$ as $t\rightarrow \infty$, we have
\begin{equation}\label{eqa:last-result-1}
P'_{\text{e-cum}}(k)\propto \frac{(2r)^{\tau -t}}{2r-1}=\frac{1}{2r-1}k^{-h(r)\ln (2r)}.
\end{equation}
where $\tau=t-h(r)\ln k$ and $h(r)=\frac{2r}{\ln (2r)}+\frac{1}{\ln (2r+m_v)}$. Clearly, $P'_{\text{e-cum}}(k)$ obeys the power law distribution and is independent of $p=(1-p_r)(1+p_a)$.

Notice that $P_{\text{e-cum}}(k)$ is equivalent to $P'_{\text{e-cum}}(k)$ by Eq.(\ref{eqa:edge-cumulative-theorem}) and Eq.(\ref{eqa:last-result-1}), also $P_{\text{e-cum}}(k)\propto 2rP'_{\text{e-cum}}(k)$. Because both models $N(t)$ and $N'(t)$ have the same growth operation and the same selection of bound-edges which leads to there are similar giant components in them. So, combining with (\ref{eqa:c3-key-form}) gives us
$$P(k)\propto \frac{(2r-1)^2}{2m_v}P_{\text{e-cum}}(k)\propto \frac{r(2r-1)^2}{m_v}P'_{\text{e-cum}}(k),$$ which enables us to conclude that the randomized EBGN-model $N'(t)$ is scale-free. The proof of Theorem \ref{them:main-thereom} is complete.\js

\end{document}